# Development of a Conceptual Framework for Knowledge Integration in Learning Newton's Third Law

Lei Bao*

Joseph C. Fritchman

The Ohio State University, Department of Physics, Columbus, Ohio 43210, USA

Abstract

Newton's third law is one of the most important concepts learned early in introductory mechanics courses; however, ample studies have documented a wide range of students' misconceptions and fragmented understandings of this concept that are difficult to change through traditional instruction. This research develops a conceptual framework model to investigate students' understanding of Newton's third law through the knowledge integration perspective. The conceptual framework is established with a central idea emphasizing forces as the outcomes of an interaction instead of using the common action-reaction language. Additional studies to apply this conceptual framework in assessment and instruction will follow.

* Corresponding author
  bao.15@osu.edu

## I. Introduction

A primary goal in introductory science education is for students to develop deep understandings of essential scientific concepts yet many students fail to achieve this after traditional instruction [1–5]. As shown by research, students may perform well on typical textbook problems with familiar contexts; however, when faced with unfamiliar contexts or problems requiring deeper understanding, they are likely to rely on pattern matching and memorized equations [2,3,6,7]. These tendencies exhibit the known characteristics of novice knowledge structures where knowledge is often locally clustered with links connecting familiar contexts [8–13]. This form of knowledge organization leads to novices employing problem-solving strategies that focus on memorized processes and solutions cued by surface features [10,14,15], which constrains novices' applications of a concept to contexts similar to those encountered through normal coursework, leaving them unable to transfer in novel situations. Meanwhile, experts' knowledge structures appear as integrated, hierarchically arranged networks built around a few core principles [12,16,17]. Their coherent knowledge organization enlists robust, far-reaching links connecting all elements, including surface features and elements deep in the abstract domain [8,10–13], and enables meaningful thinking which allows experts the ability to apply concepts across different domains and unfamiliar contexts [11,14,18].

Therefore, instruction with a goal of transitioning novices to experts should focus on developing coherent knowledge structures by fostering student abilities to build connections among new and existing ideas, which is a key process emphasized in the knowledge integration perspective of learning and instruction [17,19–22]. Practically, instruction aimed for knowledge integration focuses on the process of establishing organization within a student's knowledge structure through anchoring the structure around a central idea that serves as a core conceptual node for establishing a fully connected hierarchical network of knowledge. In the knowledge integration perspective, expert-level learners are able to use the central idea to solve problems with a wide range of contexts by connecting surface features and principles to the central idea and by determining the optimal strategies for applying the concept [17,19,20].

As an operational tool to explicitly model students' knowledge structures and measure the levels of knowledge integration achieved, the conceptual framework model was developed in previous studies [21–24], which can provide a modeling framework to illustrate the differences in knowledge structures between novices and experts by eliciting the existing connections that give rise to the range of students' alternative conceptions. Within a conceptual framework, a learner's ideas and connections are activated by contextual features. Experts link the activated ideas and related conceptual components to form specific reasoning pathways through the central idea, which extend to a fully integrated knowledge structure. The expert approach links a wide range of situations to the central idea, which can meaningfully and efficiently address complex problems in different contexts. Novices, however, often bypass the central idea and form locally memorized links between equations or algorithms and problems' surface-level features. The novice approach may produce quick and correct results in some limited cases, but it often fails to transfer to problem solving in novel situations.

Once the conceptual framework for a concept is established, it can be used as the basis for developing assessment tools, which make emphasis on probing students' knowledge structures and explicitly mapping their conceptual pathways to reveal their levels of knowledge integration. The assessment results can then inform instruction to emphasize specific conceptual pathways and connections during teaching to promote knowledge integration.

Recent studies developed conceptual frameworks for a number of physics topics, which have guided the creation of assessment questions that probe students' reasoning pathways and levels of knowledge integration [21,23,24]. These assessments contain a mixture of typical questions, which can be solved with memorization strategies, and atypical questions, which require integrated thinking involving the central idea. Usually, typical questions are designed with familiar contexts students encounter in coursework, while atypical questions are often designed with deep level conceptual understandings and using unfamiliar contexts not commonly seen during coursework. By comparing students' performances on typical and atypical questions, students' levels of knowledge integration can be empirically modeled and determined [21,24]. The assessments, combined with the underlying conceptual framework and results of student interviews, were shown to reveal students' reasoning pathways within their knowledge structures [21,23,24].

Furthermore, conceptual frameworks can aid in designing teaching interventions to improve knowledge integration by developing the essential missing links in students' knowledge structures. Ample research has demonstrated the importance of using knowledge-integration approaches in instruction to help forming integrated knowledge structures and achieving deep conceptual understanding [17, 25]. In the recent study on the conceptual framework of Force and Motion [23], an intervention, which explicitly introduced the central idea while connecting it directly to applications in problem solving, was shown to be effective in promoting knowledge integration.

Following the lead of prior studies of conceptual framework in Force and Motion, Momentum, and Light Interference [21,23,24], this research examines student understanding on Newton's third law, which is a fundamental concept in physics. In the literature, there exist a large number of studies focusing on identification of the misconceptions and difficulties related to Newton's third law [e.g. 26–38]. However, studies have shown that many misconceptions are still prevalent after instruction and may even be exacerbated, in part, due to possible misleading representations in textbooks and lectures [30,31]. Building upon the previous studies, this research develops a conceptual framework model for Newton's third law to analyze student difficulties through the knowledge integration perspective.

## II. Development of Conceptual Framework for Newton's Third Law

The most critical component of a conceptual framework is the central idea of a concept, which is identified based on experts' views and provides the core explanatory mechanisms and premises for establishing the fundamental and causal relations underpinning the concept. Additionally, related contextual variables, interactive relations, and reasoning processes are identified to form the nodes and connections of the conceptual framework. Through these processes, student difficulties in understanding the concept are carefully reviewed so that the related elements and reasoning pathways existing within students' knowledge structures can be well represented with the conceptual framework.

### A. Difficulties with Newton's Third Law

Research in physics education over the last several decades have documented extensive student difficulties and naive beliefs or misconceptions with Newtonian mechanics, which are persistent even after instruction [26,28–30,39–42]. In particular, studies on Newton's third law (N3L) reveal a range of naive beliefs, even among trained educators [28,29,31,34,35]. Many of these beliefs exist prior to entering a physics classroom as students tend to form generalizations about how the world works from an early age. Over time, these conceptions are reinforced continually and become core components of the students' knowledge structures, which can strongly resist changes to the scientific conceptions [40,42].

Most studies on student understanding of N3L focus on identifying students' misconceptions and assessing how students respond to new instruction [26–36,43]. Common examples of misconceptions include the application of a dominance principle where the faster, more massive, or acting object applies a greater force than the other object, difficulty identifying forces of a N3L pair, thinking action and reaction forces must balance each other, thinking forces are properties of objects rather than results of an interaction, and assigning causal relationships to objects and forces. Each of these points towards a fundamental misunderstanding of the nature of forces, which seem to persist even after instruction [31].

Additionally, these misconceptions exist in comparable situations absent of formal physics descriptions. Studies on psychology students revealed, most notably, a launching effect or causal asymmetry similar to the dominance principle students apply in physics [37,38,44–48], in which the observers universally attributed cause to the "acting" or "dominant" object. Further research has shown that causal asymmetry is prevalent in large portions of the general population, even in early childhood development and has been measured with children less than a year old [49–52]. Thus, by the time students encounter Newton's laws, they have years of observations reinforcing their intuitive physics misconceptions. Absent of formal physics descriptions, this demonstrates an existing bias that students may possess prior to, or concurrent with, physics learning. Ideally, the bias should be minimized through the learning of N3L. However, after instruction, students still demonstrated sensitivity to irrelevant information, such as whether kinetic energy is conserved, and systematic biases, such as believing objects with higher initial speed or mass will apply stronger forces [47,53–57].

Remedying these misconceptions is further hindered by the methods of teaching and language used. Common textbook and verbal explanations of N3L include statements such as "*for every action there is an equal and opposite reaction,*" which have been noted as flawed for at least the past eight decades [30,31]. The language itself may imply the existence of a cause (the action) and effect (the reaction) and does not emphasize the role of interaction. Textbooks have generally improved their discussions of N3L with statements such as "*We can now recognize force as an interaction between objects rather than as some 'thing' with an independent existence of its own*" [e.g. 58], but this and similar explanations are often glossed over during lecture and are not a focal point of instruction. Furthermore, most textbooks still rely on the action-reaction language which may directly lead to a belief in causality and further difficulties associated with the dominance principle.

From the literature and review of current instruction on N3L, it appears that the N3L concept is intuitively difficult for students who may hold many naturally developed misconceptions. Meanwhile, the presentation in the traditional curriculum also lacks the necessary emphasis on the core mechanism of the concept, and the use of action-reaction language can be inherently misleading. Therefore, it is important to establish a conceptual framework model for N3L to aid assessment and instruction by emphasizing the correct central idea of the N3L concept.

## B. Building the Conceptual Framework

From the commonly quoted action-reaction description of N3L, it is apparent that this definition lacks a clear description of the mechanistic nature of N3L or its basic properties, and can often lead to student difficulties [28,30,31]. The action-reaction language itself implies a time separation between action and reaction and can often be interpreted as a causal relation. However, this aspect is rarely, if ever, explicitly discussed in teaching, which leaves students to potentially draw their own conclusions on causality in N3L interactions based on colloquial interpretations of the words used rather than a concrete understanding of causality in a physics sense. Despite the issues, textbooks still commonly use a similar, yet more complete version: e.g. "*Every force occurs as one member of an action-reaction pair of forces*" [58]. Although explanations in the text have emphasized that these forces only occur in an interaction, the concept of interaction is not further repeated and enhanced in the text. Instead, the text proceeds to focus on the action-reaction and equal and opposite aspects of the interaction forces. For example, Knight [58] explicitly mentions the need for an interaction pair of forces yet still calls them an action-reaction pair. In addition, there is no explanation on the reason for using the term interaction, and the text makes very little additional emphasis on applying this idea in problem solving. The typical focus of most examples and homework is solely on the equal and opposite aspect.

Additionally, students, who are unable to recognize that N3L must involve interactions between two objects, are then often unable to identify which forces constitute a pair of interaction (or action-reaction) forces. For example, many students focus on a memorized equal-and-opposite rule where students choose any pair of forces which happen to be equal in magnitude and opposite in direction and calls them the interaction pair. In situations such as a box on a horizontal surface, this rule is often applied by students to the weight and the normal force exerted on the box. This type of understanding reveals a fragmented knowledge structure relying on memorized rules. In summary, the traditional definition of N3L emphasizes features of the interaction forces but lacks the mechanistic explanation of the origin of the forces, leading to the focus on the "action and reaction" and "equal and opposite" language. This understanding lacks the deeper conceptual foundation which is the key issue to be addressed by the conceptual framework model with the central idea of N3L.

Following reviews of expert views, student difficulties, and concerns in current N3L teaching, the central idea for N3L is identified, which states: "*An interaction between two entities can lead to a pair of symmetric (i.e., equal and opposite) interaction forces being observed.*" Here, it is important to note the unidirectional order of this definition: the result of the interaction is the symmetric forces and not the other way around. The reverse pathway is the commonly memorized rule that uses the "equal and opposite" feature to determine interaction pairs. Reasoning with the central idea then leads to the following elaborated properties which can be readily derived:

- Forces are observed outcomes of an interaction between two objects. As a result, the forces always occur in pairs acting on two interacting objects respectively and can be called a pair of interaction forces.

- Interaction forces must occur at the same time.
- Neither of the two forces causes the other force; the origin of a pair of interaction forces is the interaction.
- Interaction forces must both be the same type of forces (i.e., a pair of gravitational forces, frictional forces, normal forces, etc.)
- Interaction forces are symmetric such that they have an equal magnitude but act in opposite directions.

Building off the central idea to include different operational rules and reasoning processes from experts and novices, the conceptual framework of N3L is developed and shown in Figure 1. The top layer contains the central idea with links to the basic properties. The bottom layer contains contextual features and variables, which are commonly used as part of an N3L problem. These include surface details about the objects involved such as mass, size, velocity, or acceleration. Additionally, the context of interaction typically includes collision, push, force at a distance, or an object at rest on top of another object. The middle layers contain the intermediate reasoning processes and operational rules and procedures, which connect to the central idea in experts' knowledge structures or are locally linked through incorrect reasoning between contexts and responses by novice students. The common applications of student reasoning, such as defining the action and reaction forces based on the dominance principle or focusing on the equal and opposite properties of N3L, are often carried out by students with local links between contexts and responses through some of the naïve type intermediate processes. The conceptual framework forms the backbone for analyzing how learners reason using N3L by combining these layers and the task goals, with arrows connecting different contextual, conceptual, and outcome components representing the possible reasoning pathways of learners. Solid arrows represent experts' conceptual pathways, while the dashed-line arrows represent pathways of novices.

To achieve a deep understanding of N3L, students must be able to identify which forces are interaction forces; on which objects each of these forces act; the magnitude, type, direction, and timing of interaction forces; and what causes these forces. For experts, the central idea serves as a central node that connects to all the different properties and reasoning processes, forming an integrated knowledge structure. Therefore, when solving N3L questions, experts recognize and activate the central idea as the guiding principle in their analysis to identify relevant variables and problem-solving approaches. On the other hand, novices make weak, local connections between the different layers, forming fragmented knowledge structures. They tend to rely on memorized solutions or matching context variables with equations based the surface features of problems without deep understanding.

In N3L problems, students often encounter scenarios where one of the two interacting forces has a dominant surface feature (such as a larger mass or a higher velocity), which is irrelevant to the properties of the interaction forces based on the N3L central idea. However, novices lack the understanding of the central idea and often place focus on the dominant surface features, which lead to difficulties in answering these questions, especially when comparing magnitudes of forces [e.g. 28,37]. The incorrect reasoning on magnitude can be mapped in the conceptual framework with a pathway from the related surface features through the intermediate dominance-based reasoning to the task outcome, which in this case is the thinking that the dominant object applies a larger force than the secondary object. Often a similar difficulty occurs when students must determine causality, in which case the dominant features are often used to determine the causal relation between the interaction forces. If the students have encountered the scenario previously, they may have a memorized rule within their intermediate reasoning and processes, which can be activated and lead to the memorized solution without further processing.

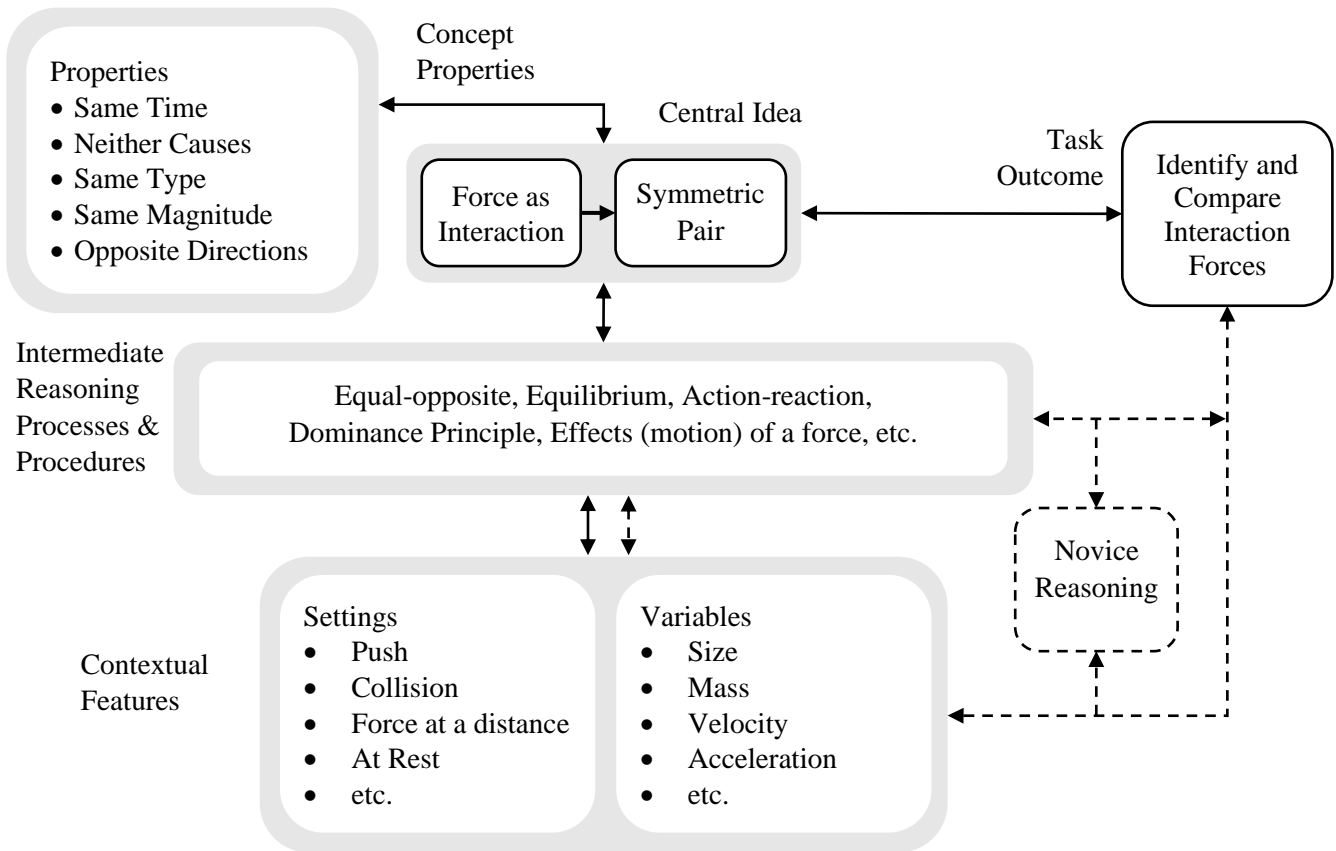

Figure 1: Conceptual Framework for Newton's Third Law. Solid arrows represent experts' conceptual pathways while dashed-line arrows represent novices' reasoning.

Additional student difficulties in N3L occur when students encounter two forces which are equal and opposite but not necessarily an interaction pair [e.g. 31,59]. This often manifests itself with students applying an equal-and-opposite rule: because two forces are equal and opposite, then they must be the action and reaction forces in N3L. Students tend to connect any given forces to this rule. For example, in the case of a stationary box in a horizontal surface, the gravitational force and the normal force applied on the box are indeed equal and opposite. As a result, students are likely to consider these two forces as an interaction pair. This type of student reasoning reveals the typical pathway of connecting surface features to a memorized rule without distinction, which then leads to incorrect outcomes. This equal-and-opposite rule is also related to the equilibrium rule that many students hold [31], where students believe that N3L requires an object to be in equilibrium. If students perceive that an object will move in a scenario, then they may assume that N3L does not apply. The conceptual framework can map this type of pathways from surface features of the question to an intermediate reasoning focusing on what motion occurs after the forces.

The novice students' pathways generally focus on the levels of contexts and specific rules in the conceptual framework, with minimal connections to the central idea. Commonly, the interaction idea is completely absent from these students' reasoning. As students advance to intermediate and expert levels, they develop more integrated knowledge structures, expanding connections to elements of the central idea. The conceptual framework can then aid in showing these differences between novice, intermediate, and expert students by comparing their knowledge structures, conceptual pathways, and problem-solving behaviors.

### C. Levels of knowledge integration within student knowledge structures

Using the conceptual framework, student misconceptions and difficulties documented in the literature can be represented and interpreted with thinking pathways of specific learning dynamics and states for analysis of students' levels of knowledge integration. In previous research, a number of developmental levels that encompass the range of student knowledge states were identified through large scale testing and interview studies on light interference and momentum [21,24]. Differences between the levels were found to be directly related to performance and reasoning on typical and atypical questions used in the assessments. Summaries of the three levels of students and their problem-solving behaviors in N3L are discussed below:

*Novice level*: The knowledge structures of novice students are typically fragmented with only local connections among surface features as the means to solve problems, leading to poor performance on both typical and atypical questions. These students base their answers on intuitive understanding of the real world and memorization of learned examples and often use contextual variables to cue their memory of likely answers without meaningful reasoning connecting to other related conceptual components or ideas. These novices can then only correctly answer a limited set of familiar questions, which they previously encountered. Some of these students struggle on even simple typical questions because they have not established either the central idea or memorized links between typical questions and results. These students often directly link surface features to their responses without meaningful reasoning or are relegated to guessing the answer when no memorized examples are applicable.

When working with typical contexts, such as questions asking students to compare the magnitude of forces in an interaction, novice students generally rely on previously memorized results or their intuitively developed naïve understandings such as those based on the dominance principle [43,60,61]. These students often focus on dominant surfaces features, such as greater mass or velocity, which students directly link to the outcome that the dominant object applies the greater force and/or causes the forces. However, when students apply a dominance principle, they likely do not link the scenario to the basic concept of N3L and fail to make connections to the N3L central idea.

On the other hand, when working with atypical questions, such as asking students to identify pairs of interaction forces, the novice students often demonstrate further weaknesses and fragmentation within their knowledge structures. These students typically place an overreliance on the equal-and-opposite rule where any two forces that are equal in magnitude and opposite in direction are likely identified as an interaction pair [31]. This rule is easily memorized and can be directly related to common statements of N3L. Many of the difficulties that novice students encounter with N3L can be traced back to this memorized rule. Students at this level simply do not have a connected knowledge structure built to reason these types of questions.

Additionally, these students do not connect other important properties of N3L to their knowledge structures. Frequently the conceptual components of time, type, and cause may be missing because they are never explicitly discussed in traditional instruction. Students who do relate the cause component to N3L may directly connect it to the action or dominant features and attribute such features as the cause of the interaction forces. These naïve views are often strongly embedded within students' experiences, which can lead to substantial difficulties in learning N3L.

*Intermediate level:* Students at this level can engage in a deeper and moderately more extended level of reasoning to develop more connected understanding with the contextual variables than the novice students; however, these students still tend to rely on memorized examples and procedures to aid their problem solving but to a lesser extent than the novice students. The increased integration within their knowledge structures allows students to think about the central idea in limited common textbook-like problems. However, with only weak understanding of the central idea, students often fail on the atypical questions. Students at this level usually exhibit diverse, rich behaviors that can lead to multiple sublevels.

In the N3L conceptual framework, students generally at this level would have greater success when asked to compare magnitudes of forces in an interaction. In a range of question scenarios, they would be able to directly link the problem to the central idea without heavy reliance on memorization or focus on surface features. Weaker students at this level link the surface features directly to the "symmetric forces" property, while more advanced students link first to the central idea (i.e., first to "interaction pair" and then to "symmetric forces") before reaching their final responses.

While reasoning in typical questions is improved, the lack of a solid understanding of the central idea often leads to wide-ranging difficulties on the atypical questions. Some of these students would attempt to use elements of the central idea but often apply them inconsistently. Many of these students still rely on the memorized equal-and-opposite rule. However, the more advanced students at this level reveal less reliance on the equal-and-opposite rule and begin to follow thought processes similar to those of the experts, especially on typical questions. This can result in a smaller gap between performances on typical and atypical questions.

As will be shown in a later section of this paper, intermediate students still do not necessarily integrate time, type, and cause into their N3L knowledge structures, but they generally perform well on time and type questions. However, most of these students do not completely understand causality in interactions, and causal reasoning is still often linked directly to dominant surface features within their knowledge structures. Overall, the intermediate students have more developed and integrated knowledge structures than novices. This allows a more consistent application of the central idea in typical questions and begins to allow the use of the central idea in limited atypical questions. Further improvement in reasoning on atypical questions would allow students to approach expert-like understanding, as described next.

*Expert-like*: Students with expert-like understanding can apply the central idea when answering both typical and atypical questions, signifying a robust, well-networked knowledge structure. This allows them to relate contextual variables to the central idea, along with many intermediate processes and related concepts, to form a comprehensive web of connections that can address a wide range of familiar and novel contexts.

For N3L, this means that these students can recognize that forces occur as the result of interaction between two objects and that these forces are observed to be symmetric and occur simultaneously. Furthermore, they recognize that the interaction forces are always the same type and occur on two distinct but interacting objects with the same magnitude and opposite directions, and neither force can be the cause of the other force. This allows the student to use the central idea in virtually any scenario. Being able to apply the central idea of a concept uniformly across multiple typical and atypical contexts is the hallmark of an integrated knowledge structure and diverges from the novices' fragmented structures where reasoning pathways are only locally connected within limited contexts and with little transferability.

In summary, these three levels show a general progression of knowledge integration and reveal how common misconceptions manifest themselves using the N3L conceptual framework. Assigning students to these levels would usually require careful interviews to determine reasoning patterns and then matching elements of these patterns to the conceptual framework. Additional work on applying the conceptual framework to develop assessment and instruction will be discussed in the follow up studies.

**Acknowledgement**

The research is supported in part by the National Science Foundation Award DUE-1712238. Any opinions, findings, and conclusions or recommendations expressed in this paper are those of the authors and do not necessarily reflect the views of the funding agencies.